\documentclass[twocolumn,flushbottom]{revtex4-1}
\usepackage{graphicx}   
\usepackage{multirow}	

\newcommand{\pt}[1]{\phantom{#1}} 
\newcommand{\extravspace}{\vphantom{$^{\textnormal{N}}$}} 
\newcommand{\iso}[1]{$^{#1}$Ne} 
\newcommand{\lra}{\leftrightarrow} 
\newcommand{\tns}[1]{\textnormal{\scriptsize{#1}}}

\begin{document}
\title{\Large Magneto-optical trapping of bosonic and fermionic neon isotopes and their mixtures: isotope shift of the ${^3P_2} \lra {^3D_3}$ transition and hyperfine constants of the ${^3D_3}$ state of \iso{21}}

%
\author{Thomas Feldker}
\author{Jan Sch\"{u}tz}
\author{Holger John}
\author{Gerhard Birkl}
\email{gerhard.birkl@physik.tu-darmstadt.de}

\affiliation{Institut f\"{u}r Angewandte Physik, Technische Universit\"{a}t Darmstadt, Schlossgartenstra\ss{}e 7, 64289 Darmstadt}
\begin{abstract}
We have magneto-optically trapped all three stable neon isotopes, including the rare \iso{21}, and all two-isotope combinations. The atoms are prepared in the metastable ${^3P_2}$ state and manipulated via laser interaction on the ${^3P_2} \lra {^3D_3}$ transition at 640.2\,nm. These cold ($T \approx 1$\,mK) and environmentally decoupled atom samples present ideal objects for precision measurements and the investigation of interactions between cold and ultracold metastable atoms. In this work, we present accurate measurements of the isotope shift of the ${^3P_2} \lra {^3D_3}$ transition and the hyperfine interaction constants of the ${^3D_3}$ state of \iso{21}. The determined isotope shifts are $(1625.9\pm0.15)$\,MHz for \iso{20} $\lra$ \iso{22}, $(855.7\pm1.0)$\,MHz for \iso{20} $\lra$ \iso{21}, and $(770.3\pm1.0)$\,MHz for \iso{21} $\lra$ \iso{22}. The obtained magnetic dipole and electric quadrupole hyperfine interaction constants are $A(^3D_3)= (-142.4\pm0.2)$\,MHz and $B(^3D_3)=(-107.7\pm1.1)$\,MHz, respectively. All measurements give a reduction of uncertainty by about one order of magnitude over previous measurements.
\end{abstract}

\maketitle

\section{Introduction}\label{sec:introduction}
Cold and ultracold atoms prepared by laser cooling and trapping techniques \cite{Metcalf:99} present an almost ideal sample of atomic systems, confined effectively in vacuum with vanishing interactions to the environment. Such ensembles have been used successfully for a variety of high-precision measurements, such as the determination of atomic and fundamental constants, or atom-atom interaction strengths.
In addition, laser prepared atoms have been further cooled to quantum degeneracy \cite{Pitaevskii:03}, opening a new field of research with high impact. 

Although initiated by work on alkali atoms, in recent years a continuously growing selection of atomic species have been investigated. A special class of atoms is given by metastable rare gas atoms, since their high internal energy makes it possible to investigate specific interaction processes 
and to apply specific detection techniques 
(for a review see \cite{CIGMA:11}). Important recent progress has led to the simultaneous trapping of different isotopes of metastable helium \cite{Stas:04}, the observation of Bose-Einstein Condensation (BEC) of metastable $^{4}$He \cite{Robert:01,DosSantos:01,Tychkov:06,Dall:07} and the observation of quantum degenerate Bose-Fermi mixtures of $^{3}$He and $^{4}$He \cite{Mcnamara:06}. In our work, we aim at extending this type of investigations to metastable neon atoms.

In neon, there exist two stable bosonic isotopes, \iso{20} (natural abundance: 90.48\%) and \iso{22} (9.25\%), and one stable fermionic isotope, \iso{21}, which is rather rare (0.27\%) \cite{Rosman}. The bosonic isotopes have no nuclear spin, whereas \iso{21} has a nuclear spin of $I=3/2$ causing a hyperfine structure of the atomic states (see Fig.~\ref{fig:sketch_levels}). The atoms are excited to the metastable ${^3P_2}$ state by electron bombardment. This state has a lifetime of 14.73\,s which has been measured with high precision in a laser-cooled sample \cite{Zinner}. For laser manipulation, the closed ${^3P_2} \lra {^3D_3}$ transition is used. Although expected to be more demanding than for metastable helium, Bose-Einstein condensation of metastable neon is pursued in our work.

In this paper, we report on the laser cooling and trapping of all stable neon isotopes \cite{Shimizu} and, for the first time, of all two-isotope combinations in a magneto-optical trap (MOT). We present the first measurements of the isotope shift of the ${^3P_2} \lra {^3D_3}$ (pair-coupling notation: $^2P_{3/2}\,3s[3/2]_2 \lra {}^2P_{3/2}\,3p[5/2]_3$) transition at 640.2\,nm and the hyperfine structure of the ${^3D_3}$ level for \iso{21} performed on a laser-cooled neon sample. All measurements give a reduction of uncertainty by about one order of magnitude over previous measurements.

\section{Multi-isotope trapping}\label{sec:trapping}
The experimental apparatus is essentially the same as used in our previous work \cite{Zinner,Spoden}: a beam of neon atoms is excited to the metastable ${^3P_2}$ state in a dc discharge and collimated using curved light fields. The atoms are then decelerated in a Zeeman slower and trapped in a MOT. The different laser fields for trapping, cooling, and detecting the atoms are derived from a dye laser using several acousto-optic frequency shifters. 

\begin{figure}
  \centering
  \includegraphics[width=0.6\linewidth]{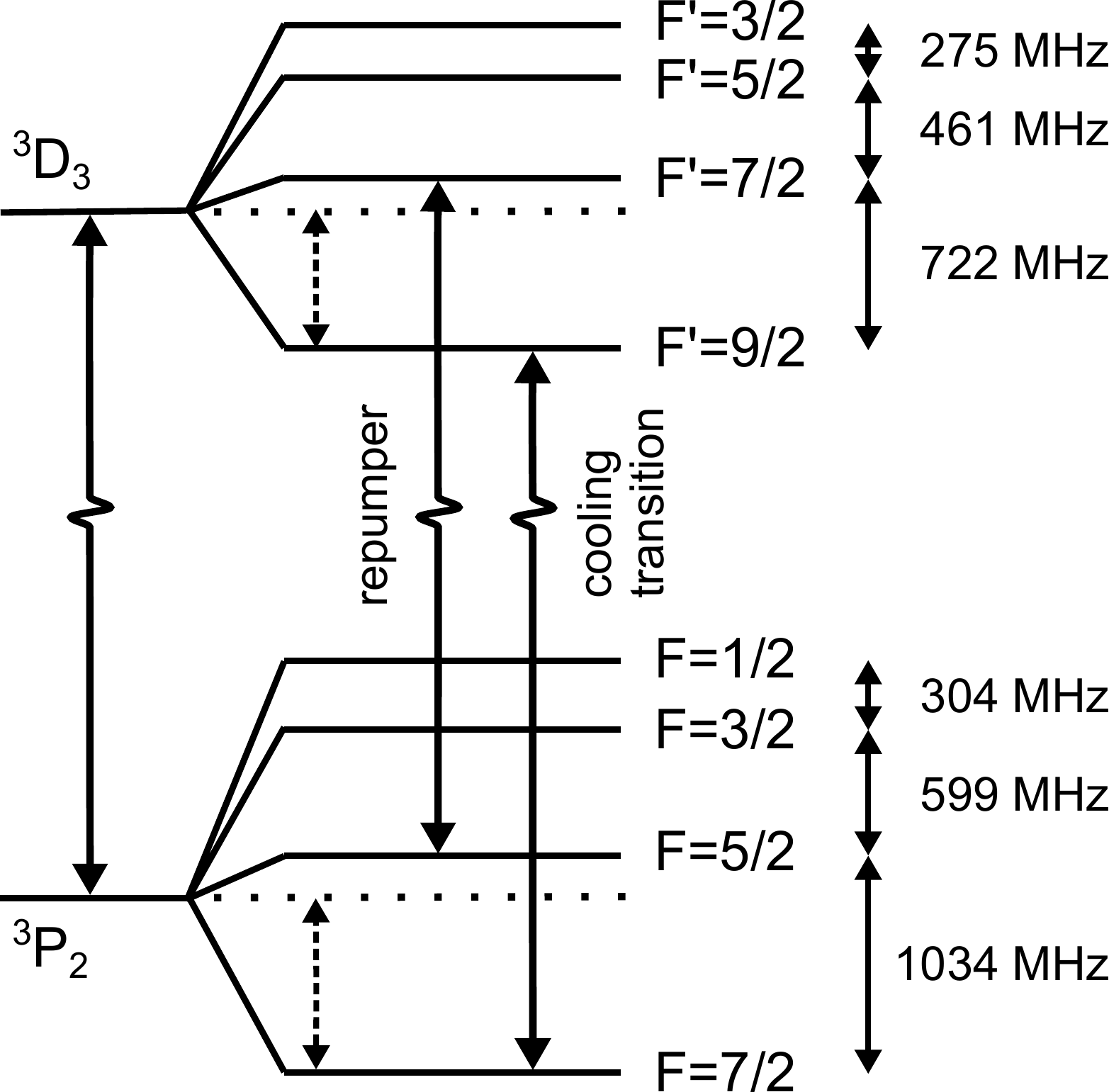}
  \caption{Hyperfine structure of the $^3P_2$ and $^3D_3$ states of \iso{21}. The atoms are trapped in the $F=7/2$ state and cooled using the transition $F=7/2\lra F'=9/2$. Repumping light is needed to transfer the atoms back form the $F=5/2$ to the $F=7/2$ state. The dashed arrows represent the hyperfine shifts that have to be known in order to obtain the nominal isotope shift of \iso{21}. The bosonic isotopes \iso{20} and \iso{22} have no hyperfine structure and no repumping light is needed.}
  \label{fig:sketch_levels}
\end{figure}

In agreement with Shimizu \textit{et al.}~\cite{Shimizu}, we find that additional laser frequencies are needed in order to trap \iso{21}. The hyperfine structure of the $^3P_2 \lra {^3D_3}$ transition is shown in Fig.~\ref{fig:sketch_levels}. In addition to the cooling light, we superimpose ``repumping'' light with a frequency resonant to the $F=5/2 \lra F'=7/2$ transition on all laser beams. This light pumps atoms that have been off-resonantly excited to $F'=7/2$ and spontaneously decayed to $F=5/2$ back into the cooling cycle. Without this additional light, trapping of \iso{21} atoms is not possible. The intensity of the repumping light is typically 18\% of the cooling light. To further increase the number of trapped \iso{21}, we optimized the loading process by pumping atoms excited to the $F=3/2$ or $F=1/2$ state in the discharge source into the cooling cycle by an additional laser beam in the optical collimation zone between discharge source and Zeeman slower. This beam has a frequency tuned between the ones of the closely spaced $F=3/2\lra F'=5/2$ and $F=1/2\lra F'=3/2$ transitions and typically has an intensity of 20\% of the main beam used for collimation. This increases the number of trapped \iso{21} atoms by about 25\%. 

The light of an additional dye laser is available for trapping a second neon isotope. By combining the light of both lasers, all two-isotope combinations can be trapped in fast succession, alternately, or simultaneously. The number of trapped atoms is determined using absorption imaging. In case of two-isotope MOTs the absorption images are taken using light resonant with the cooling transition of only one isotope. Since the isotope shift of the cooling transition is several hundreds of MHz, the presence of another isotope does not influence the imaging. Typical atom numbers are presented in Table~\ref{tab:atom_numbers}. In two-isotope traps, the atom number is about 30\% less than in the single isotope cases. The temperature is typically about 1\,mK in all cases. The number of \iso{20} and \iso{22} atoms is limited by inelastic collisions between trapped atoms \cite{Spoden,Kuppens:02,Matherson:08} while the number of \iso{21} atoms is limited by the maximum available loading duration of 0.7\,s in our setup. 

The frequency of the \iso{22} Zeeman slowing beam is only 45\,MHz detuned from the \iso{21} $F=7/2 \lra F'=7/2$ transition and, thus, this beam pumps \iso{21} atoms out of the cooling cycle. Therefore, in order to trap \iso{21} in combination with \iso{22}, we increase the repumper power and reduce the loading duration of \iso{22}. 
  
\begin{table}
  \centering
  \begin{tabular}{cc|cc}
    \hline
    \extravspace
    Isotope & Atoms $(10^6)$ & Isotopes &  Atoms $(10^6)$ \\
    \hline
    \extravspace
    \iso{20} &      600 & \iso{20} + \iso{22} & 400 + 150\\
    \iso{21} & \pt{00}3 & \iso{20} + \iso{21} & 400 + \pt{00}2\\
    \iso{22} &      200 & \iso{22} + \iso{21} & 100 + \pt{00}2\\
    \hline
  \end{tabular}
  \caption{Typical numbers of trapped atoms in single-isotope (left) and two-isotope MOTs (right). The temperature is typically 1\,mK in all cases.}
  \label{tab:atom_numbers}
\end{table}

\section{Isotope shift of the cooling transition}\label{sec:measurements}
We used these atom samples to determine the isotope shift of the laser-cooling transition with high precision. Since our approach allows for a very precise control of experimental conditions and an almost complete elimination of Doppler and collisional shifts and broadenings, our measurements are more accurate than earlier measurements using gas discharge cells \cite{Julien,Guthohrlein,Basar} or atomic beams \cite{Odintsov,Konz}. Two techniques, one based on absorption imaging and one based on imaging of the MOT fluorescence \cite{Walhout}, are used and compared to each other.

In order to determine the nominal isotope shift of the ${^3P_2} \lra {^3D_3}$ transition for \iso{21}, also the hyperfine structure of the involved energy levels has to be known. The hyperfine interaction constants of the metastable $^3P_2$ state have already been measured with high accuracy \cite{Grosof}. An accurate measurement of the $^3D_3$ hyperfine constants is presented in Section~\ref{sec:hfs} and the final results of the resulting isotope shift determination are presented in Section~\ref{sec:results}.

\subsection{Absorption measurement}\label{sec:abs}
\begin{figure*}
  \centering
  \rotatebox{90}{\hspace*{0.4cm} Absorption (arb.u.)}
  \hspace*{0.1cm}  
  \begin{minipage}[t]{0.27\textwidth}
    \centering
    \includegraphics[height=0.75\textwidth]{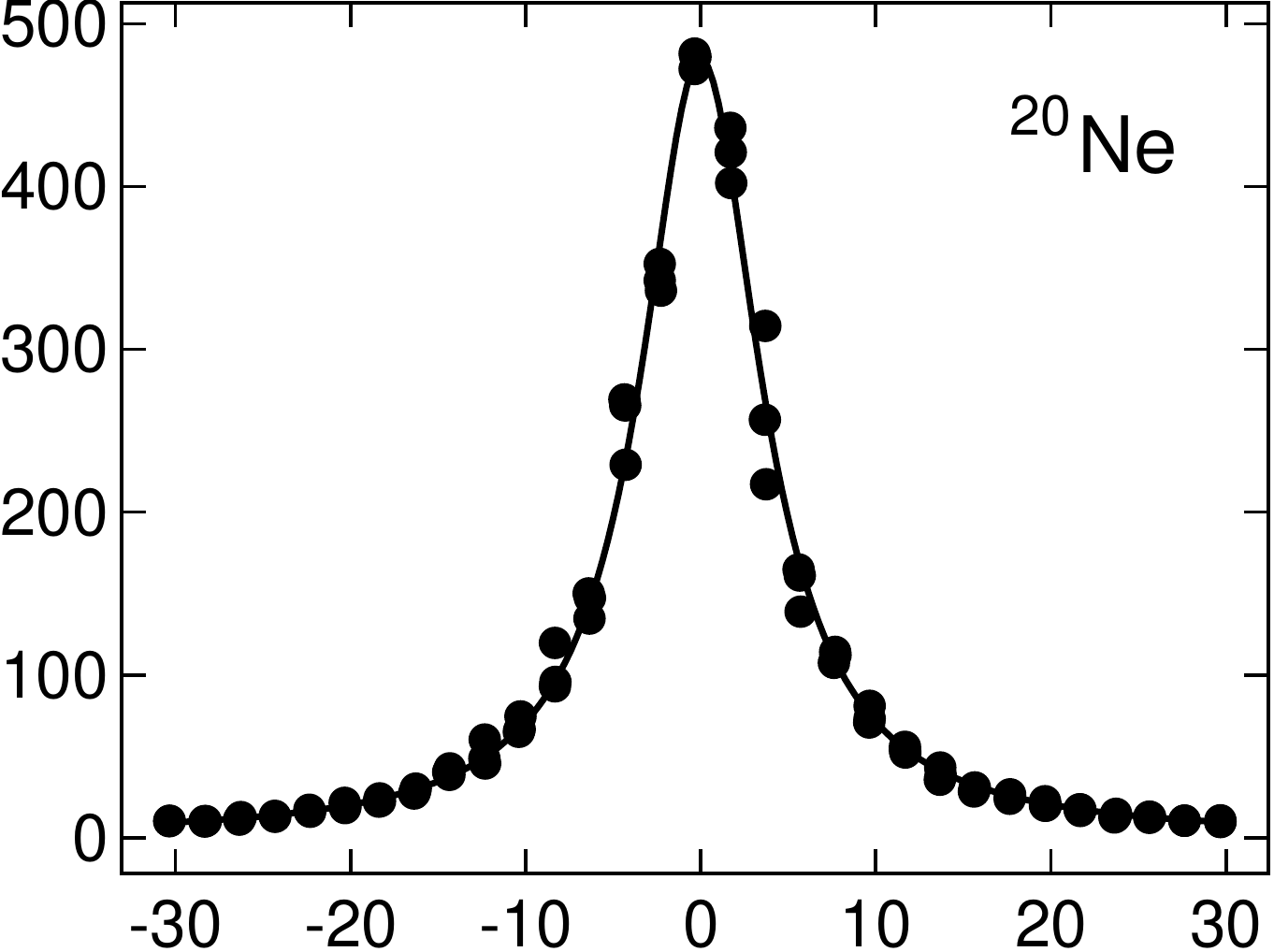}\\
	$\hspace*{0.5cm} \nu-\nu_{20}$\,(MHz)  	
  \end{minipage}
  \hspace*{0.5cm}
  \begin{minipage}[t]{0.27\textwidth}
    \centering
    \includegraphics[height=0.75\textwidth]{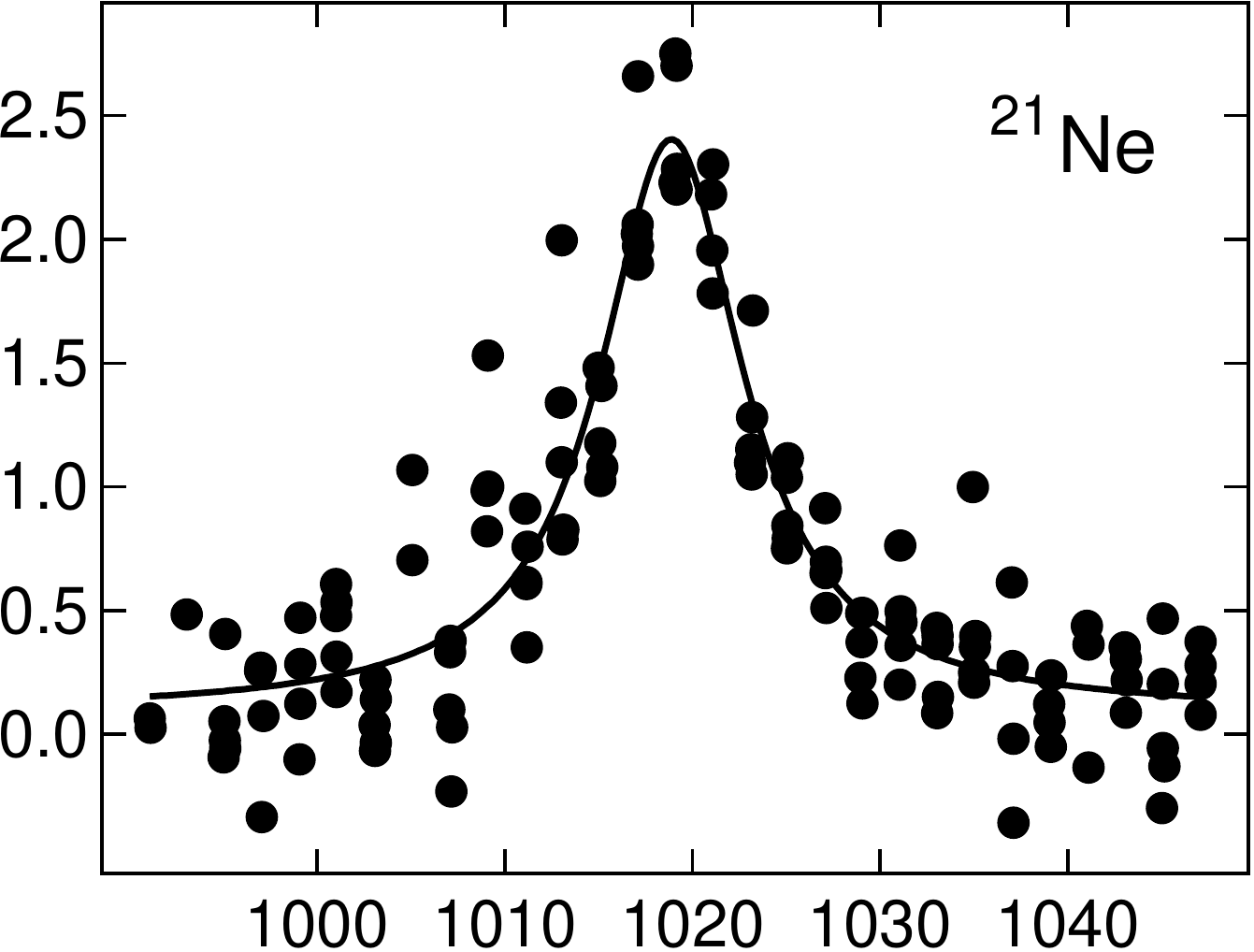}\\
	$\hspace*{0.5cm} \nu-\nu_{20}$\,(MHz)  	
  \end{minipage}
  \hspace*{0.5cm}
  \begin{minipage}[t]{0.27\textwidth}
    \centering
    \includegraphics[height=0.75\textwidth]{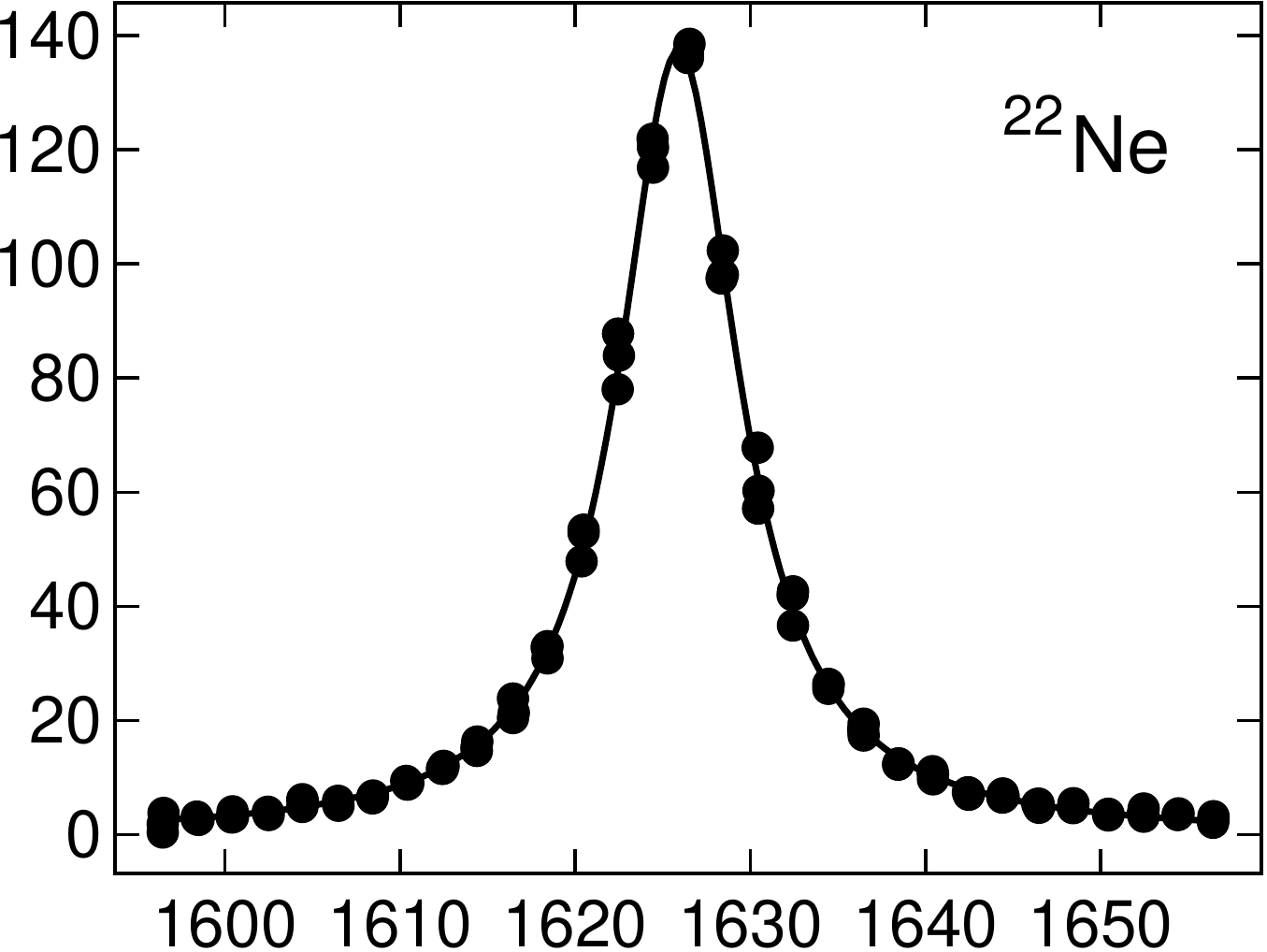}\\
	$\hspace*{0.5cm} \nu-\nu_{20}$\,(MHz)  	
  \end{minipage}    
  \caption{Absorption lines of the \iso{20}, \iso{21}, and \iso{22} cooling transitions and corresponding least square fits of Lorentzian lines. Frequencies are given relative to the \iso{20} center frequency $\nu_{20}$. The signal-to-noise ratio for \iso{21} is inferior to \iso{20} and \iso{22} due to the much lower number of trapped atoms. Each data point represents an individual absorption image. The values on resonance equal the number of trapped atoms in millions.}
  \label{fig:absorption}
\end{figure*}
In a first set of experiments, the isotope shift was determined by exploring the frequency dependence of absorption imaging. Direct measurements have been performed for the two-isotope combinations \iso{20} + \iso{22} and \iso{20} + \iso{21}, using both dye lasers which were stabilized independently to the cooling transition of each respective isotope by saturated absorption in an ac discharge cell. In each experimental run, the frequency offset between the two lasers was determined by measuring the beat frequency of two superimposed laser beams with a spectrum analyzer in order to eliminate fluctuations and drifts of the stabilization. Both isotopes were trapped alternately in direct succession to minimize effects of frequency or atom number drifts during the experimental runs, but only one isotope was trapped at a time to avoid inter-isotopic influences. 

The resulting absorption lines are shown in Fig.~\ref{fig:absorption} together with Lorentzian fits. For each data point, the frequency of the absorption beam was tuned to the designated value by an acousto-optic frequency shifter and, after the MOT beams and MOT magnetic field had been turned off, an absorption image was taken. Comparing the center frequencies of the Lorentzian lines yields the isotope shift of the cooling transition. The results are given in Table~\ref{tab:results_is}. No direct measurements where performed for the \iso{21} $\lra$ \iso{22} isotope shift in absorption, but the respective shift of the cooling transition can be calculated from the other two measurements to be $(607.0\pm0.47)$\,MHz.

The frequency shift between the cooling transitions of the bosonic isotopes was measured with high precision, whereas the measurement involving \iso{21} is less precise due to the lower number of trapped \iso{21} atoms. The uncertainties taken into account are summarized in Table~\ref{tab:errors_is}. The statistical errors of the center frequencies were calculated for each absorption line individually accounting for fit errors as well as atom number drifts. Laser drifts were deduced from the temporal drift of the beat frequency. Long-term stability and accuracy of the spectrum analyzer were tested using a rubidium frequency standard.

\begin{table}
  \centering
  \begin{tabular}{c|c}
    \hline
      &  Frequency (MHz) \\
    \hline
    Absorption measurement & \\
    \iso{20} $\lra$ \iso{22} & 1625.9\,$\pm$\,0.15 \\
    \iso{20} $\lra$ \iso{21} & 1018.9\,$\pm$\,0.45 \\
    \hline
    Fluorescence measurement & \\
    \iso{20} $\lra$ \iso{22} & 1626.0\,$\pm$\,0.22 \\
    \iso{20} $\lra$ \iso{21} & 1018.8\,$\pm$\,0.25 \\
    \iso{21} $\lra$ \iso{22} & \pt{1}607.2\,$\pm$\,0.25 \\
    \hline
  \end{tabular}
  \caption{Isotope shift of the cooling transition $^3P_2 \lra {^3D_3}$ (\iso{21}: $F=7/2\lra F'=9/2$) determined by absorption and fluorescence imaging. No direct absorption measurement was performed for the \iso{21} $\lra$ \iso{22} shift.}
  \label{tab:results_is}
\end{table}

\begin{table}
  \centering
  \begin{tabular}{c|ccc}
    \hline
         & \multirow{2}{*}{Statistics} & \multirow{2}{*}{Laser drift} & Spectrum \\
         &                              &                             & analyzer \\
    \hline
	Absorption\\
	\iso{20} $\lra$ \iso{22} & 0.14\,MHz & 0.02\,MHz & 0.05\,MHz \\
	\iso{20} $\lra$ \iso{21} & 0.42\,MHz & 0.15\,MHz & 0.05\,MHz \\
	\hline
	Fluorescence\\
	\iso{20} $\lra$ \iso{22} & 0.07\,MHz & 0.20\,MHz & 0.05\,MHz \\
	\iso{20} $\lra$ \iso{21} & 0.15\,MHz & 0.20\,MHz & 0.05\,MHz \\
	\iso{21} $\lra$ \iso{22} & 0.15\,MHz & 0.20\,MHz & 0.05\,MHz \\
    \hline
  \end{tabular}  
  \caption{Contributions to the uncertainty of the cooling transition isotope shift for absorption and fluorescence measurements.}
  \label{tab:errors_is}
\end{table}

\subsection{Fluorescence measurement}\label{sec:fluor}
\begin{figure}
  \centering
  \rotatebox{90}{\hspace*{0.9cm} Fluorescence (arb.u.)}
  \hspace*{0.1cm}  
  \begin{minipage}[t]{0.35\textwidth}
    \centering
    \includegraphics[width=\textwidth]{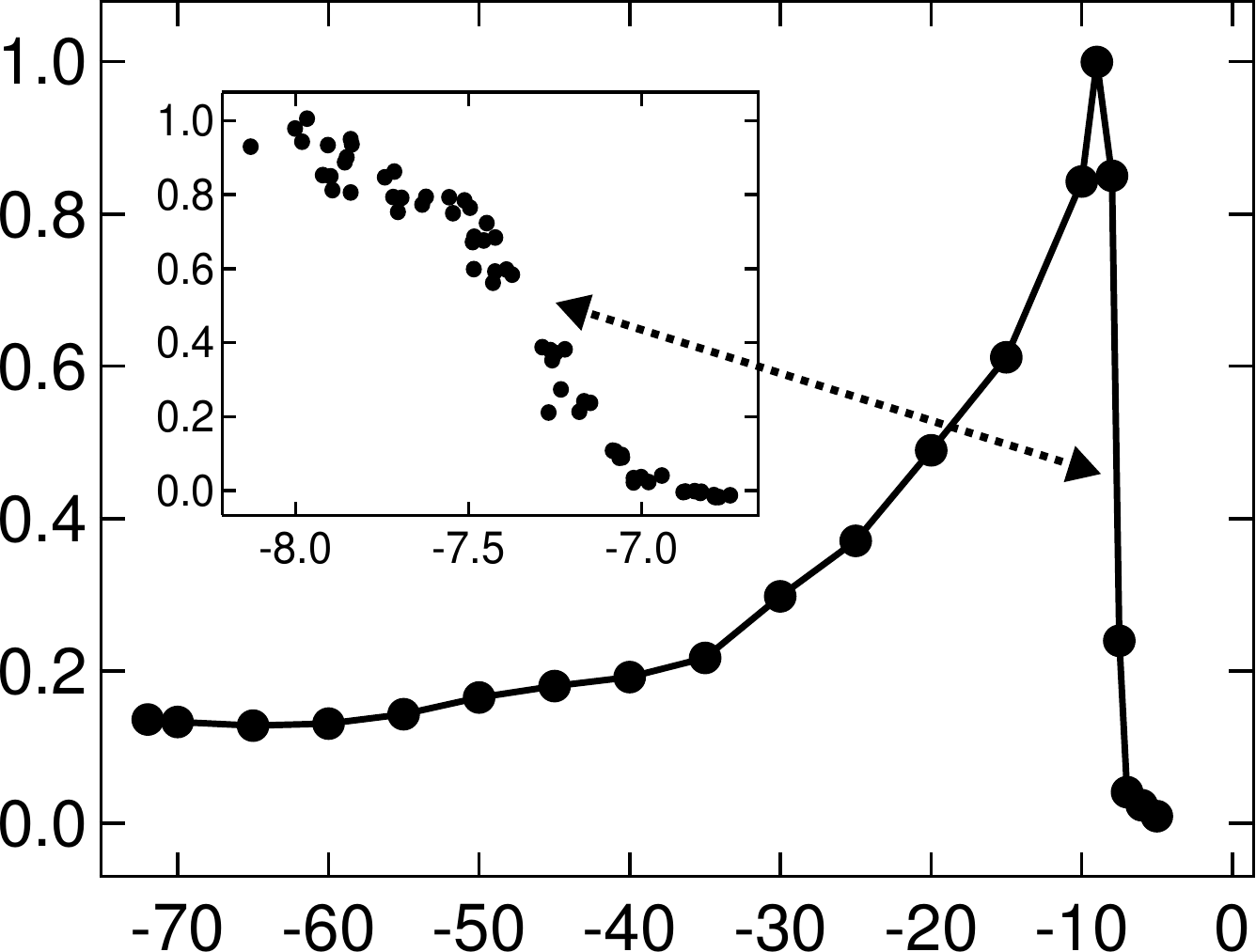}\\
	$\hspace*{0.5cm} \nu-\nu_{20}$\,(MHz)  	
  \end{minipage}  
  \caption{Fluorescence signal of a \iso{20} MOT as a function of detuning. The fluorescence decreases rapidly close to resonance due to the ineffectiveness of laser cooling. The falling edge, which is shown in greater detail in the inset, is about 500\,kHz wide.}
  \label{fig:fluorescence_signal}
\end{figure}
\begin{figure*}
  \centering
  \rotatebox{90}{\hspace*{0.4cm} Fluorescence (arb.u.)}
  \hspace*{0.1cm}  
  \begin{minipage}[t]{0.27\textwidth}
    \centering
    \includegraphics[height=0.75\textwidth]{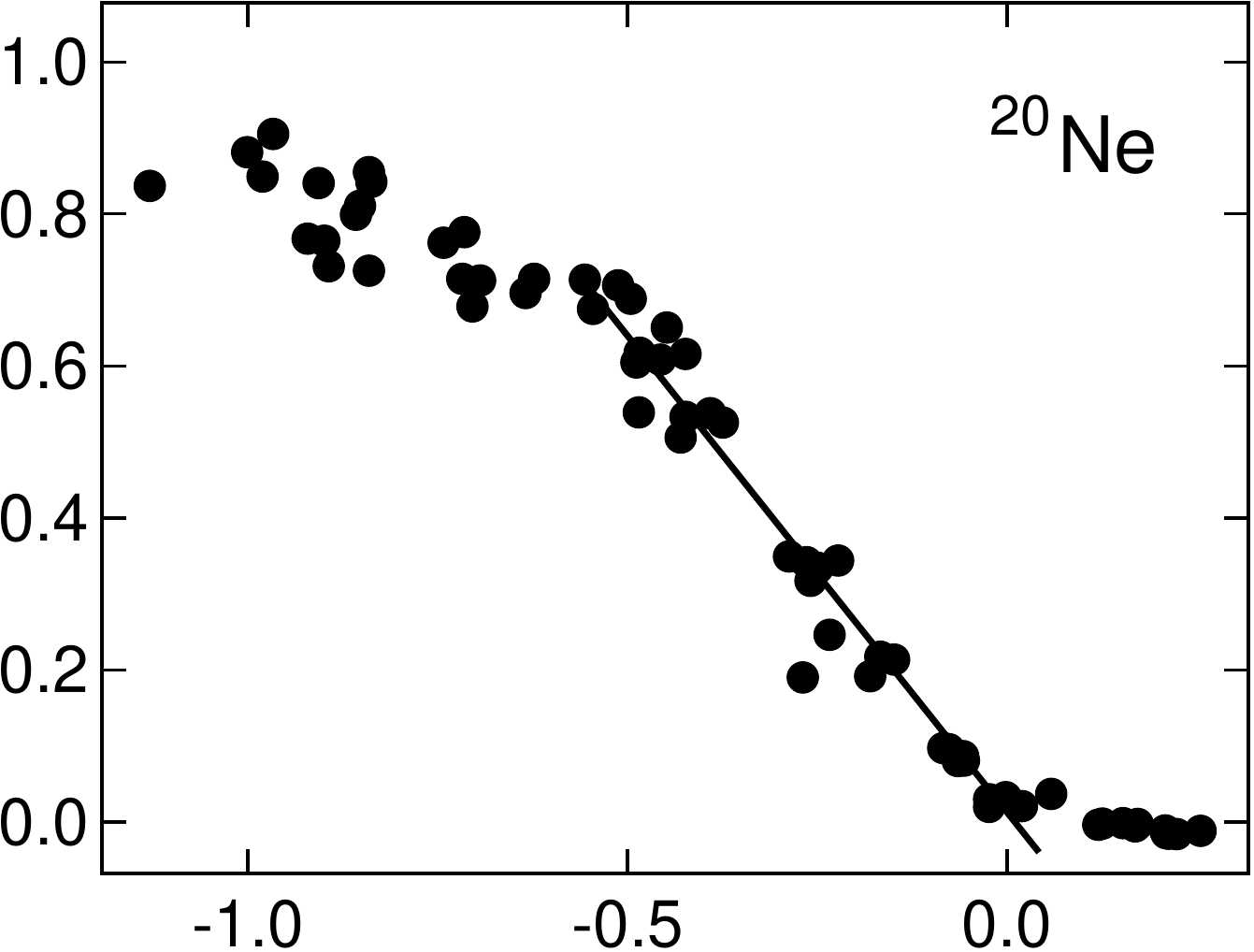}\\
	$\hspace*{0.5cm} \nu-\nu'_{20}$\,(MHz)  	
  \end{minipage}
  \hspace*{0.5cm}
  \begin{minipage}[t]{0.27\textwidth}
    \centering
    \includegraphics[height=0.75\textwidth]{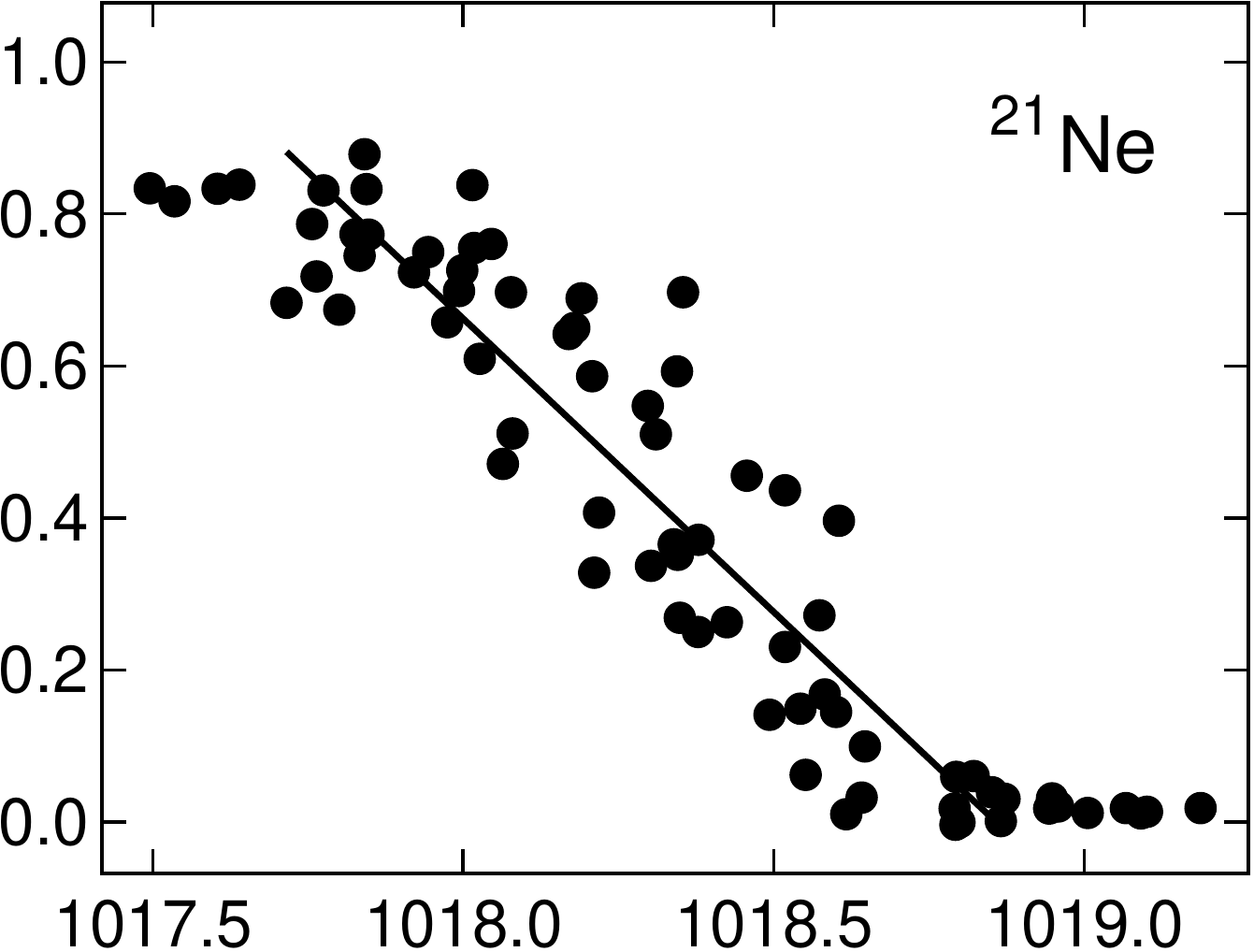}\\
	$\hspace*{0.5cm} \nu-\nu'_{20}$\,(MHz)  	
  \end{minipage}
  \hspace*{0.5cm}
  \begin{minipage}[t]{0.27\textwidth}
    \centering
    \includegraphics[height=0.75\textwidth]{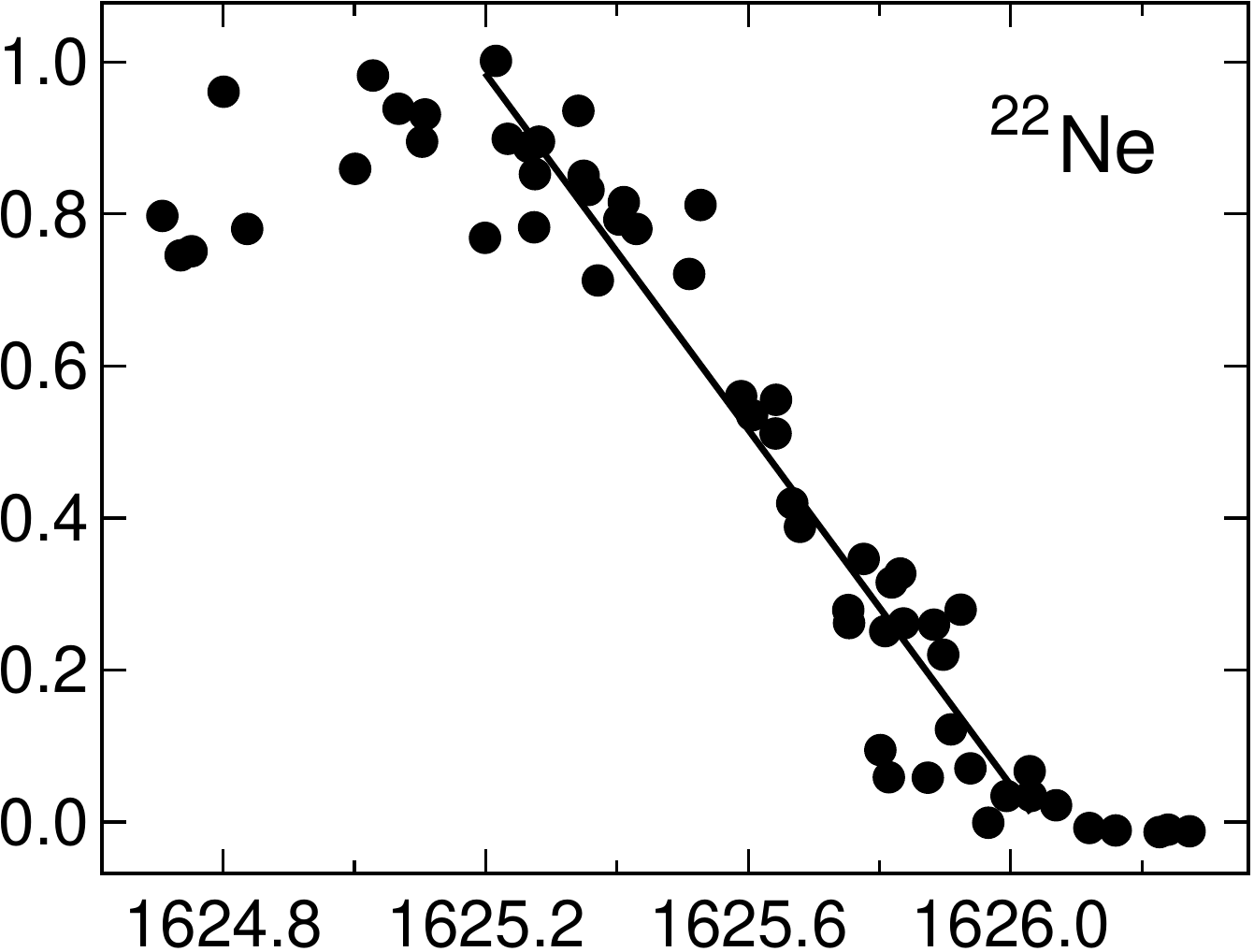}\\
	$\hspace*{0.5cm} \nu-\nu'_{20}$\,(MHz)  	
  \end{minipage}   
  \caption{Fluorescence signals of \iso{20}, \iso{21}, and \iso{22} MOTs as a function of laser frequency. Isotope shifts are calculated based on the zero crossings of the linear fits to the falling edge of the fluorescence signal. Frequencies are given relative to the \iso{20} zero crossing $\nu'_{20}$. About $2\times10^6$ atoms were trapped for each isotope. Each data point represents an individual fluorescence image.}
  \label{fig:fluorescence}
\end{figure*} 
A second determination of the cooling transition isotope shift was performed by measuring the frequency dependence of the MOT fluorescence for each isotope. When the detuning of the MOT laser beams is varied from red- ($\nu_{\tns{Laser}}-\nu_{\tns{Atom}} < 0$) to blue-detuning ($\nu_{\tns{Laser}}-\nu_{\tns{Atom}} > 0$), the MOT fluorescence rapidly decreases close to resonance \cite{Walhout}. As can be seen in Fig.~\ref{fig:fluorescence_signal}, the falling edge of the fluorescence signal is only about 500\,kHz wide as compared to the 8.2\,MHz natural line width of the transition. Since the fluorescence signal is well pronounced even for small and dilute atomic samples, this method is especially suitable for measurements including the rare \iso{21}.

The measurement was performed for the three isotopes consecutively using the same laser. The second dye laser was locked to the \iso{20} cooling transition as a stable frequency reference. In each experimental run, the MOT was loaded under constant conditions. To minimize density dependent effects, a similar number of atoms was trapped for all isotopes by blocking the optical collimation and reducing the loading time for \iso{20} and \iso{22}. After loading, the frequency of the MOT beams was tuned to the designated value using an acousto-optic frequency shifter and, after a waiting time of 50\,ms, the MOT fluorescence was recorded with a CCD camera for 50\,ms. In case of $^{21}$Ne the repumper was switched off during the fluorescence measurement.

The resulting fluorescence signals are shown in Fig.~\ref{fig:fluorescence}. A linear function is fitted to the falling edge of the signals. Isotope shifts are determined using the frequency values of the zero crossings of these fits. As can be seen in Fig.~\ref{fig:fluorescence_signal}, these frequencies $\nu'$ are about 7\,MHz below the resonance frequency $\nu$ for each isotope. Since we are only interested in frequency differences between isotopes, this constant offset cancels for the isotope shift determination. The values for the isotope shift of the cooling transition measured in this way are given in lower part of Table~\ref{tab:results_is}. The sources of uncertainties which have to be taken into account are the same as for the absorption measurements with the respective values presented in the lower part of Table~\ref{tab:errors_is}. The statistical errors, especially for \iso{21}, are smaller than for the absorption measurements due to the better signal-to-noise ratio for low atom numbers and the narrow falling edge of the fluorescence signal. The cooling transition isotope shifts determined with both methods agree well within the error margins.

\section{Hyperfine interaction constants}\label{sec:hfs}
To obtain the nominal isotope shift of \iso{21} not only the shift of the cooling transition but also the hyperfine splitting of the $^3P_2$ and $^3D_3$ states have to be known (cf. Fig.~\ref{fig:sketch_levels}). With magnetic dipole and electric quadrupole interaction constants $A$ and $B$, the splitting of the states can be calculated according to
\begin{equation}\label{eq:hfs}
  \Delta\nu_{\textnormal{\tiny{HFS}}}(F)= A\,\frac{K}{2} + B\,\frac{3/4\,K(K+1)-I(I+1)\,J(J+1)}{2I(2I-1)\,J(2J-1)}\,,
\end{equation}
with
\begin{equation}
  K = F(F+1)-I(I+1)-J(J+1)\,,
\end{equation}
where $J$, $I$, and $F$ denote the electronic, nuclear, and total angular momentum quantum numbers. The $^3P_2$ hyperfine interaction constants have been measured with high accuracy by Grosof \textit{et al.} \cite{Grosof} as $A=(-267.68 \pm 0.03)$\,MHz and $B=(-111.55 \pm 0.1)$\,MHz. The $^3D_3$ hyperfine constants had been known only to MHz accuracy \cite{Julien} and, therefore, have been also determined in this work. 

\begin{table}
  \centering
    \begin{tabular}{c|c}
      \hline
      Hyperfine splitting & Frequency (MHz) \\
      \hline
      \extravspace
      $F'=9/2 \lra F'=7/2$ &  \pt{1}721.7\,$\pm$\,0.5 \\
      $F'=9/2 \lra F'=5/2$ &       1182.5\,$\pm$\,0.5 \\
      \hline
    \end{tabular}
  \caption{Hyperfine splitting of the $^3D_3$ state of \iso{21}.}
  \label{tab:results_hfs}
\end{table}

\begin{table}
  \centering
    \begin{tabular}{c|cc}
      \hline
      \extravspace
       & $F'=9/2 \lra 7/2$ 	     & $F'=9/2 \lra 5/2$ \\
      \hline
      \extravspace
      Statistics        & 0.40\,MHz & 0.36\,MHz \\
      Laser drift       & 0.20\,MHz & 0.20\,MHz \\
      Spectrum analyzer & 0.05\,MHz & 0.05\,MHz \\
      Magnetic field    & 0.23\,MHz & 0.25\,MHz \\
      \hline
    \end{tabular}
  \caption{Contributions to the uncertainty of the measurement of the hyperfine interaction constants of \iso{21}.}
  \label{tab:errors_hfs}
\end{table}

\begin{table*}[!hbt]
  \centering
    \begin{tabular}{c|c|c|c|c|c|c}
    \hline
     & This work & Ref.\,\cite{Julien} & Ref.\,\cite{Guthohrlein} & Ref.\,\cite{Basar} & Ref.\,\cite{Odintsov} &  Ref.\,\cite{Konz} \\
     & (MHz) & (MHz) & (MHz) & (MHz) & (MHz)& (MHz)\\
    \hline
    HFS constants            &                 &                      &                &            &            & \\
    $A(^3D_3)$               & $-142.4\pm0.2$  & $-142.5\pm 1.5$      &                &            &            & \\
    $B(^3D_3)$               & $-107.7\pm1.1$  & $  -113\pm 8$        &                &            &            & \\
    \hline
    Isotope shift            &                 &                      &                &            &            & \\
    \iso{20} $\lra$ \iso{22} & $1625.9\pm0.15$ & $1630\pm3$           & $1631.2\pm5.0$ & $1629.5\pm1.0$ & $1628\pm3$ & $1632\pm3$ \\
    \iso{20} $\lra$ \iso{21} & $ 855.7\pm1.0 $ & $\phantom{1}856\pm7$ &                &            &            & \\
    \iso{21} $\lra$ \iso{22} & $ 770.3\pm1.0 $ &                      &                &            &            & \\
    \hline
    \end{tabular}        
  \caption{Measured isotope shifts of the $^3P_2 \lra{} ^3D_3$ transition of \iso{20}, \iso{21}, and \iso{22} and of the hyperfine interaction constants of the $^3D_3$ state of \iso{21} compared to previous measurements \cite{Julien,Guthohrlein,Basar,Odintsov,Konz}.}
  \label{tab:results}
\end{table*}

\begin{figure}
  \centering
  \rotatebox{90}{\hspace*{0.0cm} Fluorescence (arb.u.)}
  \begin{minipage}[t]{0.22\textwidth}
    \centering
    \includegraphics[height=0.73\textwidth]{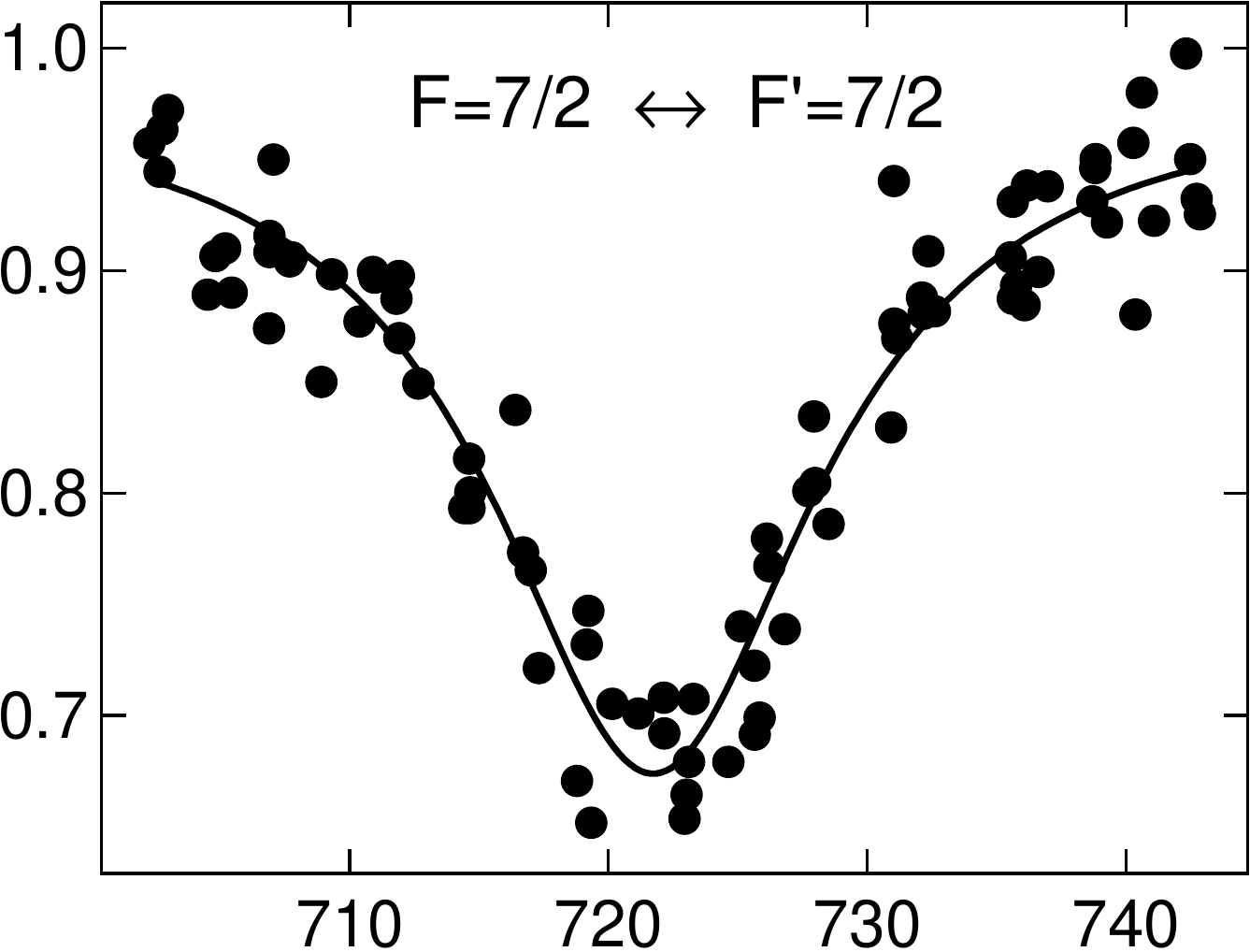}\\
	$\hspace*{0.5cm} \nu-\nu_{(7/2\lra9/2)}$\,(MHz)  	
  \end{minipage}
  \begin{minipage}[t]{0.22\textwidth}
    \centering
    \includegraphics[height=0.73\textwidth]{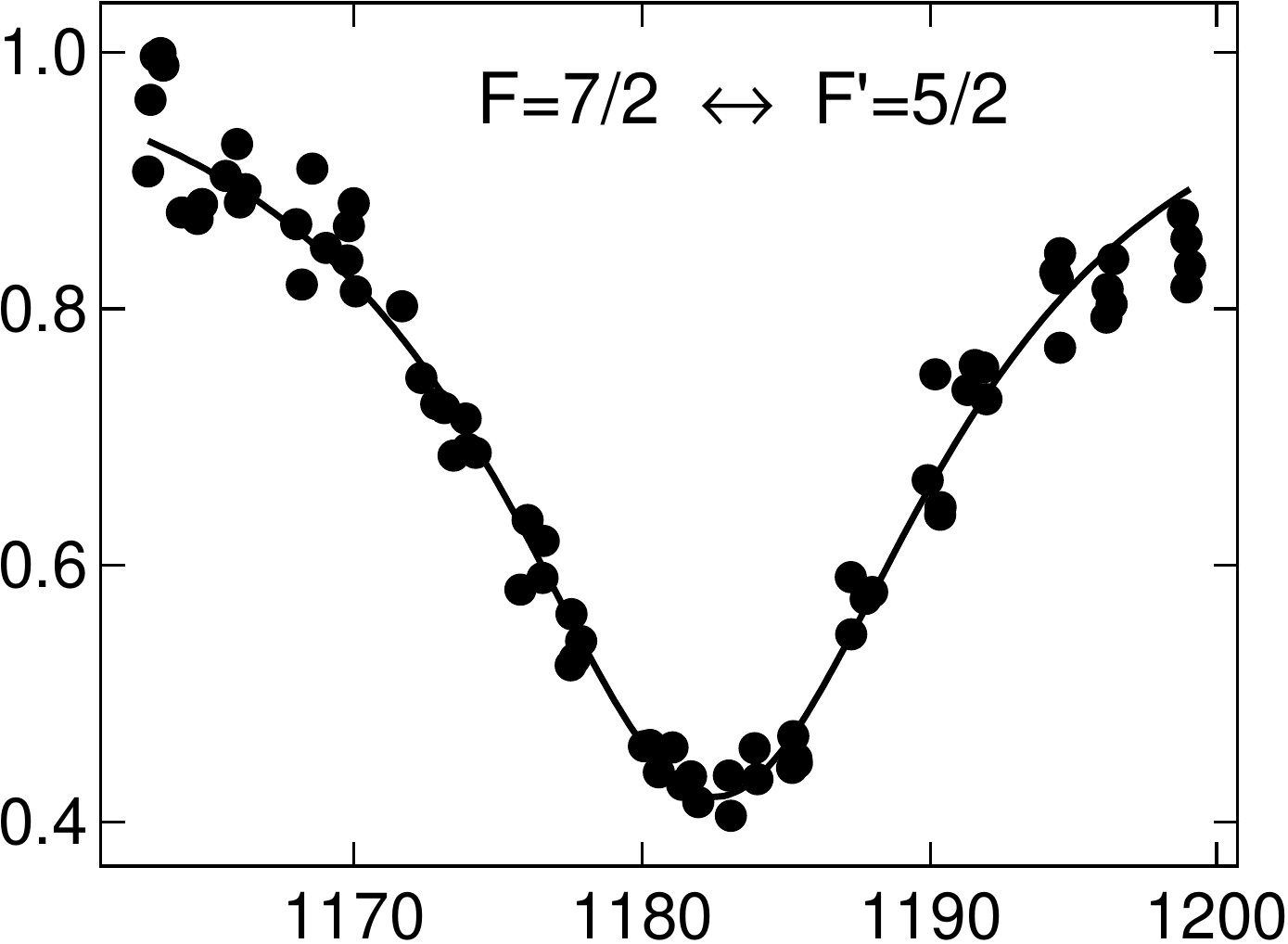}\\
	$\hspace*{0.5cm} \nu-\nu_{(7/2\lra9/2)}$\,(MHz)  	
  \end{minipage}  
  \caption{Fluorescence signal and Lorentzian fits for \iso{21} in a MOT after irradiation by a probe pulse which is scanned over the $F=7/2 \lra F'=7/2$ and $F=7/2 \lra F'=5/2$ resonances. Frequencies are given relative to the $F=7/2 \lra F'=9/2$ transition frequency.}
  \label{fig:hfs}
\end{figure}

To obtain the hyperfine interaction constants, we measured the splittings between the $F'=9/2$, 7/2, and 5/2 levels of the $^3D_3$ state. For this purpose, we measured the trap loss caused by an additional probe beam resonant to the $F=7/2 \lra F'=7/2$ or $F=7/2 \lra F'=5/2$ transition which pumps atoms out of the cooling cycle. In each experimental run about $2\times10^6$ atoms in the $F=7/2$ state were loaded into the MOT which was then switched off. After 1.5\,ms waiting time to allow for magnetic fields to vanish, the probe pulse was applied for 360\,$\mu$s and the beat frequency between probe and cooling laser light was determined using a spectrum analyzer. The MOT was then switched on again without repumping light to retrap all atoms remaining in $F=7/2$. The MOT fluorescence was measured with the CCD camera for 50\,ms in order to determine the number of the retrapped atoms. In addition, the frequency of the $F=7/2 \lra F'=9/2$ transition was measured using absorption imaging as described in Section~\ref{sec:abs}. The trap loss as a function of probe pulse frequency for the $F=7/2 \lra F'=7/2$ and $F=7/2 \lra F'=5/2$ transitions are shown in Fig.~\ref{fig:hfs}. The center frequencies were determined by least square fits of Lorentzian functions. The measured splittings are given in Table~\ref{tab:results_hfs}. 

The hyperfine interaction constants calculated according to Eq.~\ref{eq:hfs} are $A=(-142.4 \pm 0.2)$\,MHz and $B=(-107.7 \pm 1.1)$\,MHz. The sources of uncertainties are the same as described in Section~\ref{sec:abs} plus an additional uncertainty taking into account the effect of imperfect compensation of magnetic fields since transitions involving states with different g-factors have been measured. The remaining magnetic field was found to be less than 17\,$\mu$T (170\,mG) which leads to a frequency shift of 0.15\,MHz to 0.20\,MHz, depending on the transition. The resulting uncertainties caused by the magnetic field were evaluated by averaging the shifts of all possible $F,m_F$ $\lra$ $F',m'_F$ transitions assuming an equal initial distribution between states. All contributing uncertainties are summarized in Table~\ref{tab:errors_hfs}.

\section{Final results and discussion}\label{sec:results}
The final results for the nominal isotope shift of the $^3P_2 \lra{} ^3D_3$ transition and the hyperfine interaction constants of the $^3D_3$ level of \iso{21} are compiled in Table~\ref{tab:results} together with previous experimental results of Refs.~\cite{Julien,Guthohrlein,Basar,Odintsov,Konz}. Due to the smaller uncertainty, the \iso{20} $\lra$ \iso{22} isotope shift is taken from the absorption imaging results (Section~\ref{sec:abs}). The values for isotope shifts involving \iso{21} are based on the fluorescence imaging results (Section~\ref{sec:fluor}), the $^3P_2$ hyperfine interaction constants from Ref.~\cite{Grosof}, and the $^3D_3$ hyperfine constants from Section~\ref{sec:hfs}. 

Compared to previous measurements \cite{Julien,Guthohrlein,Basar,Odintsov,Konz}, our results are a factor of 7 to 10 more precise, which demonstrates the advantage of using laser cooled and trapped atoms for precision measurements.
Our and the previous results agree well within the error margins, except for the \iso{20} $\lra$ \iso{22} isotope shift which is found to be about 4\,MHz smaller in our work. While the reason for this discrepancy is unknown, our results obtained for several independent measurement runs and by using two independent methods (absorption and fluorescence method) agree well within our small error margins. We have checked our methods very carefully for additional systematic errors but could not identify any significant additional sources of uncertainty. To our knowledge, no ab initio calculations for a comparison to our results are available in the literature. The high-precision results presented here should trigger respective calculations and should help to improve the determination of the mass and charge radii of $^{17-22}$Ne including the proton-halo candidate $^{17}$Ne \cite{Geithner:08}.

We would like to thank Alexander Martin, Norbert Herschbach, and Gabriele Jenny-Deu\ss{}er for helpful discussions. This work was supported in part by the Deutsche Forschungsgemeinschaft (DFG), by the BMBF (contract numbers 06DA9019I and 06DA9020I), and by the European Science Foundation (ESF) within the CRP CIGMA of EuroQUAM. 

\bibliography{Feldker_isotopeshift}

\end{document}